\documentclass[twocolumn,letter]{jpsj3}
\usepackage{amsmath,amssymb,bm}
\usepackage{color}

\begin{document}

\title{Local Electron Correlations in a Two-dimensional Hubbard Model 
on the Penrose Lattice}%

\author{Nayuta \surname{Takemori}
\thanks{E-mail address: takemori@stat.phys.titech.ac.jp} and 
\name{Akihisa \surname{Koga}}
}
\inst{\address{Department of Physics, Tokyo Institute of Technology, 
Meguro, Tokyo 152-8551, Japan}
}
\abst{
We study electron correlations in the half-filled Hubbard model 
on two-dimensional Penrose lattice.
Applying the real-space dynamical mean-field theory to large clusters, 
we discuss  how low-temperature properties are affected 
by the quasiperiodic structure.
By calculating the double occupancy and renormalization factor at each site, 
we clarify the existence of the Mott transition.
The spatially-dependent renormalization characteristic of geometrical structure is also addressed.
}

\kword{Hubbard model, Penrose lattice, 
dynamical mean-field theory}

\maketitle
Quasiperiodic systems have attracted considerable interest 
since the discovery of quasicrystal~\cite{Shechtman84}.
One of the specific features is the existence of 
the long-range order without translational symmetry. 
This should induce interesting low-temperature properties 
in the metallic quasicrystals such as 
electric and thermal conductivities~\cite{Kimura91,Inaba02}.
The tight-binding model on the quasiperiodic lattices,
which may describe some quasicrystal compounds,
has been studied and intrinsic properties such as 
the existence of the confined state and fractal dimensions,
have been clarified~\cite{Kohmoto86,Tokihiro88}.
Recently, interesting low-temperature properties have been 
observed in the quasicrystal $\rm{Au_{51}Al_{34}Yb_{15}}$
and its approximant $\rm Au_{51}Al_{35}Yb_{14}$~\cite{Deguchi12}, 
which stimulates the further theoretical investigations~\cite{Watanabe,Popov}.
In the former compound, 
the specific heat and susceptibility exhibit power-law behavior with
a nontrivial exponent at low temperatures.
In contrast, the approximant with the periodic structure shows
conventional heavy fermion behavior.
These should suggest that electron correlations and 
quasiperiodic structure play 
a crucial role in stabilizing quantum critical behavior at low temperatures.
Therefore, it is desirable to clarify how electron correlations affect
low temperature properties in the quasiperiodic system.

\begin{figure}[htb]
\includegraphics[width=8cm]{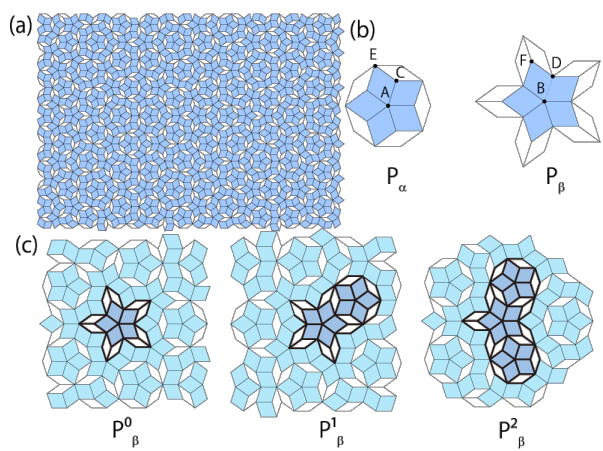}
\caption{\label{fig:f1} 
(Color online) (a) Two-dimensional Penrose lattice. 
The lattice is composed of shaded (fat) and 
open (skinny) rhombuses.
(b) Pentagon structures $\rm{P_\alpha}$ and $\rm{P_\beta}$.
We label center sites in $\rm{P_\alpha}$($\rm{P_\beta}$) as A(B) and nearest neighbor(NN) sites and next nearest neighbor(NNN) sites of center sites in $\rm{P_\alpha}$($\rm{P_\beta}$) as B(C) and D(E).
(c) Possible overlap structures ${\rm P}_\beta^i$.
}
\end{figure}

Motivated by this, we study correlated electron systems 
on the quasiperiodic lattice.
One of the simplest questions is how the introduction of the Coulomb 
interaction forms the quasiparticles and leads to the Mott transition
in the system
since coordination number and geometrical structure depend on 
the lattice sites.
To attack this fundamental problem, 
we focus on the half-filled Hubbard model on a two-dimensional Penrose lattice,
where a site is placed on each vertex of the rhombuses
[see Fig. \ref{fig:f1}(a)].
We apply the real-space dynamical mean-field theory (RDMFT)~\cite{Georges96} 
to the Hubbard model,
and discuss electron correlations in the system.
Calculating renormalization factor and double occupancy 
at each site, we study how the Coulomb interactions yield 
the spatially-dependent renormalization, 
typically close to the Mott transition point.

In this paper, 
we consider the single-band Hubbard model on the Penrose lattice,
which should be given by the following Hamiltonian
\begin{equation}
H=-t\sum_{\langle i, j \rangle} (c^{\dagger}_{i \sigma} c_{j \sigma}+ h.c.)+U \sum_{i}n_{i \uparrow}n_{i \downarrow},
\end{equation}
where $\langle i,j \rangle$ denotes the nearest neighbor site, 
$c^{\dagger}_{i \sigma} (c_{i \sigma})$ is 
a creation (annihilation) operator of an electron 
at the $i$th site with spin $\sigma(= \uparrow , \downarrow)$ 
and $n_{i \sigma}= c^{\dagger}_{i \sigma}c_{i \sigma}$.
$t$ is the transfer integral between sites and 
$U$ is the Coulomb interaction.
The Penrose lattice we treat here is bipartite, and 
its ground state is expected to be
an antiferromagnetically ordered state when $U\neq 0$.
Namely, the spontaneous staggered magnetization appears
in the Heisenberg limit ($U/t\rightarrow \infty$)~\cite{Jag07}.
In contrast, 
electron correlations in the paramagnetic state may not be trivial.
When $U=0$, there appears an interesting density of states due to
the local lattice geometry.
As shown in Fig. \ref{fig:f1}, there are many pentagon structures 
in the lattice and each 
central vertex has locally a five-fold rotational symmetry,
which leads to a number of isolated confined states~\cite{Kohmoto86}. 
When the system is half-filled, the corresponding energy is located 
at the Fermi level
and it may be difficult to treat electron correlations in the system.
This characteristic feature is common 
to the half-filled Hubbard models on 
the checkerboard~\cite{Fujimoto},
Lieb~\cite{Noda}, and pyrochlore lattices~\cite{pyrochlore}. 
In addition to the above noninteracting feature, 
the spatial dependence of the renormalization is 
highly nontrivial in the quasiperiodic system
since the coordination number ranges from 3 to 7,
and geometrical structure depend on the lattice site.

Here, we consider the Hubbard model with large clusters 
under the open boundary condition
to avoid the influences from the periodicity and edge of the system.
For this purpose,
we use the RDMFT method, where local electron correlations are taken into account.
This enables us to obtain reliable results 
if spatially extended correlations are negligible. 
In fact, the method has successfully been applied to 
correlated particle systems such
as surface~\cite{surface}, interface~\cite{interface}, 
fermionic atoms~\cite{cold1} 
 and topological insulating systems~\cite{topological}.

In RDMFT, the lattice model is mapped to the effective impurity models
dynamically connected to each "heat bath".
The lattice Green's function is then obtained via the self-consistency 
condition imposed on these impurity problems.
In the framework of RDMFT, 
the lattice Green's function ${\hat G}_{\rm lat}$ is given 
in terms of the site-diagonal self-energy $\Sigma_i$ as
\begin{eqnarray}
[{\hat G}^{-1}_{\rm{lat}}(i\omega_n)]_{ij}=-t \delta_{\langle ij \rangle} + (i \omega_n+\mu)\delta_{ij} -\Sigma_i(i \omega_n),
\end{eqnarray}
where $\mu$ is the chemical potential, $\omega_n=(2n+1)\pi T$ is 
the Matsubara frequency, and $T$ is the temperature.
The local self-energy $\Sigma_i$ and Green's function ${G}^i_{\rm imp}$
are obtained by solving the effective impurity model for the $i$th site. 
We use the hybridization-expansion continuous-time quantum Monte Carlo 
 (CTQMC)
 method~\cite{Werner} as an impurity solver.
The self-consistent loop of calculations is iterated under the condition 
$[{\hat G}_{\rm lat}( i \omega_n)]_{ii}=
G_{\rm imp}^i (i \omega_n)$.

In the paper, we treat the half-filled system, setting the chemical potential as $\mu=U/2$. To focus on the Mott transition at low temperatures, we neglect ordered phases such as the density wave and magnetically ordered states, and consider the paramagnetic solution with $\langle n_{i\sigma}\rangle=0.5$.
In order to study low-temperature properties, 
we calculate the double occupancy 
$d_i=\langle n_{i \uparrow} n_{i \downarrow} \rangle$
at $i$th site.
Furthermore, we deduce the quantity 
$z_i=\left[ 1 - {\rm Im} \Sigma_i (i\omega_0)/\omega_0  \right]^{-1}$ 
as a renormalization factor at finite temperatures.
Although this should not be the quasiparticle weight for the local density of states, it is appropriate to discuss local electron correlations and Mott transitions.
In the following, we use the hopping $t$ as the unit of the energy.
For convenience, we treat the Penrose lattice 
with the five-fold rotational symmetry, which is iteratively generated in terms of inflation-deflation rule~\cite{inflation}.

\begin{figure}[htb]
\includegraphics[width=8cm]{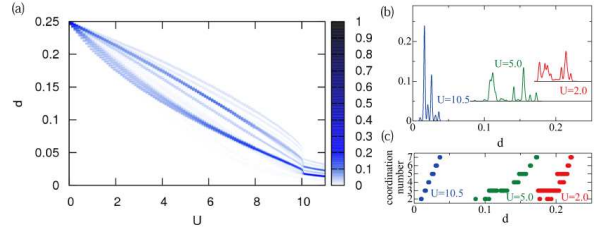}
\caption{\label{fig:density} 
(Color online) Density plot of double occupancy (a) and
its cross-section (b) 
in the half-filled Penrose-Hubbard model with 4481 sites when $T=0.05$.
(c)The distribution of the double occupancy for the coordination number of lattice sites.
The unit of energy is set to be $t$.
}
\end{figure}
We first discuss how the Mott transition occurs at low temperatures,
applying the RDMFT method to the Penrose Hubbard model with 4481 sites.
In the system with the five-fold rotational symmetry, 
there exist 444 independent sites, 
and the corresponding local quantities are, in general, 
different from each other.
Fig. \ref{fig:density}(a) shows
the distributions of the double occupancy $d_i$ at the temperature $T=0.05$.
In the noninteracting case $U=0$, the normal metallic state is realized
with $d_i=0.25$.
Introducing the interaction, the double occupancy at each site 
monotonically decreases, 
similar to the conventional Hubbard model~\cite{single}.
In a certain interacting case, it is found that the quantities range
in a certain width, as shown in Fig. \ref{fig:density}(b).
This means that the site-dependent renormalization occurs.
Further increase in the interaction yields a jump singularity 
in each curve of $d_i$ at $U=U_{c2}$,
where the first-order phase transition occurs 
to the Mott insulating state.
The transition point is obtained as $U_{c2}\sim 10.1$.
As discussed in detail later, another jump singularity in some curves appears close to the Mott transition point, which might be invisible in this scale.
When the interaction decreases from the Mott insulating state,
the phase transition occurs to the metallic state 
at $U=U_{c1}\;(\sim 9.7)$ (not shown).
As for the Mott transition at finite temperatures,
the above results are essentially the same as those 
in the conventional Hubbard model~\cite{single}.

\begin{figure}[htb]
\includegraphics[width=8cm]{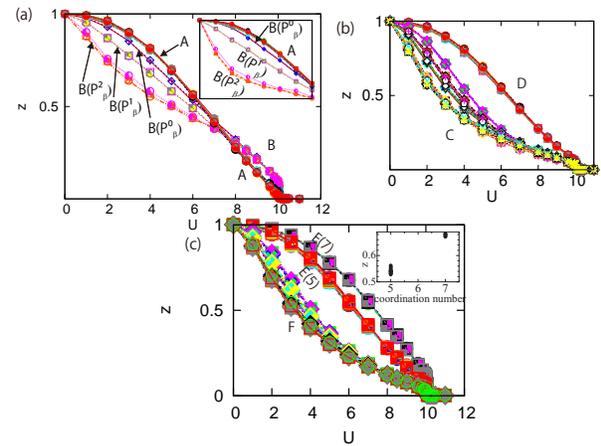}
\caption{\label{fig:zdo} 
(Color online)   
Renormalization factors for A and B sites (a), C and D sites (b), E and F sites (c)
as a function of the interaction when $T=0.05$.
Inset of (a) shows renormalization factor for A and B sites 
at a lower temperature $T=0.02$.
The unit of energy is set to be $t$.
}
\end{figure}

On the other hand, there appear interesting site-dependent renormalizations 
in the Penrose-Hubbard model.
Here, we focus on two pentagon structures $\rm{P_\alpha}$ and $\rm{P_\beta}$ 
characteristic of the Penrose lattice, as shown in Fig. \ref{fig:f1}(b).
$\rm P_\alpha$ $\rm (P_\beta)$ 
is composed of five fat tiles and five (ten) skinny tiles.
In addition to the isolated structure, 
there are the overlap structures ${\rm{P_\beta}}^i$
composed of one pentagon structure $\rm{P_\beta}$ and 
$i(=0, 1, 2)$ pentagon structures $\rm{P_\alpha}$,
as shown in Fig. \ref{fig:f1}(c).
To clarify how a local geometry affects electron correlations,
we show the renormalization factor $z_i$ in Fig. \ref{fig:zdo}. 
Although there exist 275 distinct A and B sites 
in the Penrose lattice with 4481 sites, 
Fig. \ref{fig:zdo}(a) shows that
the quantities are almost grouped into four classes, which are characterized by
A sites in $\rm{P_\alpha}$, B sites in  $\rm{P_\beta^0}$, $\rm{P_\beta^1}$, and $\rm{P_\beta^2}$.
The renormalization factors for B sites in $\rm{P^1_\beta}$ and $\rm{P^2_\beta}$ are 
strongly suppressed by the introduction of the interaction, 
compared with the others.
This may suggest that the overlap structure effectively decreases 
hopping integrals between B and D sites, 
which enhances electron correlations relatively.
Such behavior is more clearly found at lower temperatures, 
as shown in the inset of Fig. \ref{fig:zdo} (a).
When the system approaches the Mott transition point, 
different behavior appears, 
where the renormalization factors for ${\rm{P_\beta}}^i$ merge.
This may suggest that the increase of the interaction gradually decreases
the correlation length and the presence of the overlap structure
little affects electron correlations at its center site.
In this case, particles are strongly renormalized in A sites as shown in Fig. 3(a).

The difference of the renormalizations on the pentagon structures 
is also found in the NN and NNN sites,
as shown in Fig. \ref{fig:zdo}(b) and (c).
 In the weak coupling region, the renormalization factors for C sites in $\rm{P_\alpha}$ widely range, which should be grouped into some classes depending on the overlap structure and its geometry. On the other hand, the renormalization factors for D sites in $\rm{P_\beta}$ little range. This means that, even in the weak coupling region, the overlap structure little affects electron correlations for D sites
, contrast to that in center sites as well as the NNN sites. 
We note that two classes E(5) or E(7) clearly appear in the renormalization factor for E sites in $\rm{P_\alpha}$ since these sites connect to 5 or 7 NN sites, as shown in Fig. \ref{fig:zdo}(c).
On the other hand, in the strong coupling region, 
site-dependent renormalization for C sites in $\rm{P_\alpha}$ and F sites in $\rm{P_\beta}$ disappears.
These facts mean that 
the local renormalization is not directly related to the correlation length 
but depends on its pentagon structure. 
Therefore, we can say that nontrivial electron correlations appear 
in the metallic state of the system.

Beyond the Mott transition point $(U>U_{c_2})$, 
we find that the distribution of the double occupancy for all sites
is grouped into five classes, as shown in Figs. \ref{fig:density} (b) and (c).
This implies that, in the strong-coupling region, 
the local quantities at each site depend on its effective hopping.
Namely, the coordination number ranges from 3 to 7 in the Penrose lattice, and some sites with two connecting bonds appear only at the edge of the system.

\begin{figure}[htb]
\begin{center}
\includegraphics[width=6cm]{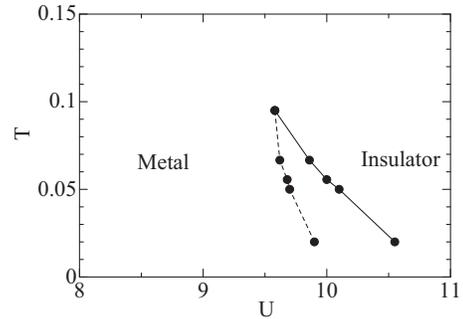}
\end{center}
\caption{\label{fig:f2} 
(Color online) Phase diagram of the half-filled Hubbard model 
on two-dimensional Penrose lattice.
The dashed and solid lines represent the phase boundaries $U_{c1}$ and $U_{c2}$,
respectively.
}
\end{figure}

By performing similar calculations, we obtain the phase diagram, 
as shown in Fig. \ref{fig:f2}.
There are two phase boundaries $U_{c1}$ and $U_{c2}$, 
where the first-order phase transition occurs between 
the metallic and Mott insulating states with decreasing (increasing)
interactions.
Increasing temperatures, these two boundaries merge at 
the critical end point around $(U/D, T/D) \sim (9.5, 0.094)$.
At higher temperatures, there are no singularities in physical quantities 
and the crossover between the metallic and insulating states occurs.
At very low-temperatures, inter-site correlations should be important,
which could not be treated in our method correctly.
It is an interesting problem how magnetic fluctuations develop, 
which is now under consideration.

\begin{figure}[htb]
\includegraphics[width=8cm]{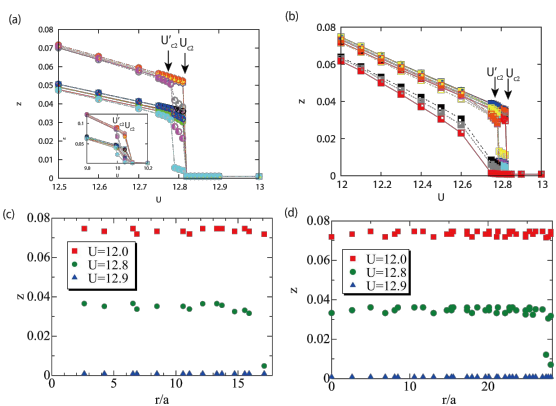}
\caption{\label{fig:f6b20IPT} (Color online) 
Renormalization factors for A sites 
in the system with 4181 (a) and 11006 (b) sites when $T=0.02$. 
Arrows represent the transition points $U_{c2}'$ and $U_{c2}$.
(c) and (d) show the renormalization factors as a function of the radius from 
the center of the system with 4181 and 11006 sites, respectively.
Squares, circles, and triangles represent 
the results of metallic, intermediate, and insulating states.
Inset of (a) shows the renormalization factors for A sites in the system with 4181 sites, which are obtained from the RDMFT and CTQMC method.
The unit of energy is set to be $t$.
}
\end{figure}

We finally discuss how low temperature properties close to the Mott transition
are affected by the open boundary conditions.
To this end, 
we use the iterative perturbation theory 
(IPT) as an impurity solver.
This method is based on the second-order perturbation theory
and is appropriate to describe the Mott transitions
in the half-filled Hubbard model.
In fact, the Mott transition point is obtained as $U_{c2}=12.8$ $(T=0.05)$, 
which is consistent with the CTQMC results. 
Furthermore, this method has an advantage in efficiently evaluating 
the self-energy of the effective impurity model 
in comparison with the CTQMC method.
This allows us to treat larger clusters systematically.
The renormalization factors for A sites 
in the system with 4181 and 11006 sites are shown in 
Figs. \ref{fig:f6b20IPT}(a) and (b). 
It is found that the renormalization factor for each A site 
ranges in the tiny width ($0.065<d_i<0.075$) when $U=12.0$, as discussed before.
However, the quantities for certain sites decrease differently 
when the system approaches the Mott transition point. 
We then find the jump singularity in the curves of the renormalization factors for some sites at $U_{c2}'$, which is slightly smaller than $U_{c2}$. These are consistent with the CTQMC results, as shown in the inset of Fig.5(a).
To clarify the nature of this phenomenon,
we show in Figs. \ref{fig:f6b20IPT}(c) and (d) 
the renormalization factors of A sites 
as functions of the distance from the center of the system.
When the system belongs to the metallic (insulating) state,
the renormalization factors are finite (almost zero) in the system. 
On the other hand, in the intermediate region $U_{c2}'<U<U_{c2}$,
the renormalization factors in the bulk are different from 
those around the edge of the system.
Therefore, this behavior should originate from the open boundary
of the system. 
Since a single Mott transition has been found in the RDMFT calculations with IPT on the square lattice under the open boundary condition ($39\times 39$ sites; not shown), we believe that this behavior is not due to our numerical technique.
Namely, in the intermediate region $U_{c_2}'<U<U_{c_2}$, the renormalization factors for edge sites are still finite and the local density of states at the Fermi level remains (not shown).
Examining the effect of open boundary condition on the other sites, and deducing the width of the edge region, 
we find that the thicknesses is independent for the system size ($d/a \sim 2$), where $a$ is the length of a side of rhombus.
Therefore, we can say that large clusters should be important to discuss
the nature of the Mott transitions in quasiperiodic systems.

In this work, we have studied the Penrose Hubbard model and have clarified that electron correlations
strongly depend on the site and its geometry.
However, it was difficult to clarify whether or not the above interesting renormalization originates from the quasiperiodic structure although we have treated the large clusters.
Furthermore, we did not find the detailed structure
e.g. self-similarity in the physical quantities.
Therefore, further detailed investigations
on strong correlations in the quasiperiodic systems are desired.

In summary, we have investigated the half-filled Hubbard model 
on the two-dimensional Penrose lattice, 
combining the RDMFT with the CTQMC method.
Computing the double occupancy and renormalization factor at each site, 
we have discussed the Mott transition at finite temperatures.
Furthermore, we have found that the quasiparticle weight strongly depends 
on the site and its geometry.
It is also interesting to study the periodic Anderson model 
on the quasiperiodic lattice 
since the valence of the ytterbium ions in the compound 
$\rm Au_{51}Al_{34}Yb_{15}$ is intermediate~\cite{intermediate}, 
which is now under consideration.

The authors would like to thank R. Peters, T. Pruschke, and S. Wessel 
for valuable discussions. 
This work was partly supported by 
the Grant-in-Aid for Scientific Research 25800193\ and 
the Global COE Program ``Nanoscience and Quantum Physics" 
from MEXT of Japan.
Part of the computations was carried out on TSUBAME2.0
at Global Scientific Information and the Computing Center
of Tokyo Institute of Technology and at the Supercomputer
Center at the Institute for Solid State Physics, University of
Tokyo.
The simulations have
been performed using some ALPS libraries~\cite{alps}.

\end{document}